\documentclass{jfm}
\usepackage{graphicx}
\usepackage{amsmath}
\usepackage{tabularx}
\usepackage{color}
\usepackage{comment}
\usepackage[colorlinks,citecolor = blue, linkcolor=red,hyperindex,CJKbookmarks]{hyperref}

\usepackage{siunitx}
\newcolumntype{Y}{>{\centering\arraybackslash}X}

\shorttitle{Solvent exchange in porous media}
\shortauthor{Y. Li, K. L. Chong, H. Bazyar, R. G. H. Lammertink and D. Lohse}

\title{Universality in microdroplet nucleation during solvent exchange in Hele-Shaw like channels}

\author{Yanshen Li,\aff{1}
\corresp{\email{yanshen.li@utwente.nl}}
  Kai Leong Chong,\aff{1}
  Hanieh Bazyar,\aff{2,3}
  Rob G. H. Lammertink,\aff{2}
 \and Detlef Lohse,\aff{1,4}
 \corresp{\email{d.lohse@utwente.nl}}
}

\affiliation{\aff{1}Physics of Fluids group, Max-Planck Center Twente for Complex Fluid Dynamics, Department of Science and Technology, Mesa+ Institute, and J. M. Burgers Centre for Fluid Dynamics, University of Twente, P.O. Box 217, 7500 AE Enschede, The Netherlands
\aff{2}Soft Matter, Fluidics and Interfaces (SFI), Department of Science and Technology, University of Twente, Enschede, The Netherlands.
\aff{3}Wetsus, European Centre of Excellence for Sustainable Water Technology, Leeuwarden, The Netherlands
\aff{4}Max Planck Institute for Dynamics and Self-Organization, 37077 G\"ottingen, Germany}

\begin{document}

\maketitle

\begin{abstract}
Micro and nanodroplets have many important applications such as in drug delivery, liquid-liquid extraction, nanomaterial synthesis and cosmetics. A commonly used method to generate a large number of micro or nanodroplets in one simple step is solvent exchange (also called nanoprecipitation), in which a good solvent of the droplet phase is displaced by a poor one, generating an oversaturation pulse that leads to droplet nucleation. Despite its crucial importance, the droplet growth resulting from the oversaturation pulse in this ternary system is still poorly understood. We experimentally and theoretically study this growth in Hele-Shaw like channels by measuring the total volume of the oil droplets that nucleates out of it. In order to prevent the oversaturated oil from exiting the channel, we decorated some of the channels with a porous region in the middle. Solvent exchange is performed with various solution compositions, flow rates and channel geometries, and the measured droplets volume is found to increase with the P\'eclet number $Pe$ with an approximate effective power law $V\propto Pe^{0.50}$. A theoretical model is developed to account for this finding. With this model we can indeed explain the $V\propto Pe^{1/2}$ scaling, including the prefactor, which can collapse all data of the ``porous'' channels onto one universal curve, irrespective of channel geometry and composition of the mixtures. Our work provides a macroscopic approach to this bottom-up method of droplet generation and may guide further studies on oversaturation and nucleation in ternary systems.

\end{abstract}

\section{Introduction}
Micro- and nanodroplets generation is of tremendous interest due to its wide range of applications in drug delivery \citep{gursoy2004self, attama2005vitro, devarajan2011nanoemulsions}, liquid-liquid extraction \citep{jain2011recent, rezaee2006determination, rezaee2010evolution,  yu2010microfluidic}, (nano)material synthesis \citep{liff2007high, kumar2008methods, duraiswamy2009droplet}, catalytic reactions \citep{shen2014oxygen, yabushita2009anion}, and cosmetics \citep{xu2005generation, lee2008generation, yeh2009using, kuehne2011highly}, etc. One way to generate microdroplets is to utilize microfluidic devices such as T-junctions \citep{yeh2009using}, flow focusing setups \citep{anna2003formation, teh2008droplet, seemann2011droplet} or co-flowing devices \citep{utada2005monodisperse, shah2008designer, serra2008microfluidic}, where monodispersed microdroplets with well-defined properties could be generated successively. All these devices \& methods utilize a top-down approach in which a liquid jet or drop is split into smaller parts. This limits the smallest droplet size which can be achieved. 

This limitation can be overcome in a bottom-up approach such as solvent exchange \citep{lou2000nanobubbles, zhang2015formation, lohse2015surface, lohse2020physicochemical}, where a large number of micro and nanodroplets is generated by nucleation out of an oversaturated solution. This method, also called nanoprecipitation or solvent shifting \citep{fessi1989nanocapsule,galindo2004physicochemical, aubry2009nanoprecipitation, lepeltier2014nanoprecipitation, hajian2015formation}, though commonly used, is much less well understood.

In solvent exchange, a good solvent of the target droplet component (the solute) is replaced by a poor solvent, where the two solvents are miscible. A typical example is an oil saturated aqueous ethanol solution being replaced by oil saturated water. Upon contact of the ethanol \& water solution, the two solvents start to mix with each other. Due to the addition of water, the solubility of oil is lowered, and the subsequent oversaturation leads to droplet nucleation and growth. The micro \& nanodroplets can nucleate in the bulk \citep{vitale2003liquid} or on a hydrophobic surface. For solvent exchange in a microchannel, it has been found that droplets nucleated in the bulk tends to migrate to and then stay in the center in a co-flowing device \citep{hajian2015formation}, where the droplet movement is controlled by solutal Marangoni flow and composition of the mixture. On the other hand, for droplets that nucleated on the surface, their average volume is found to increase with the P\'eclet number $Pe$ as $\propto h^3Pe^{3/4}$ \citep{zhang2015formation}, where the flow rate $Q$ is included in $Pe$ and $h$ the channel height. Later, the effect of flow geometry \citep{yu2017formation} and solution composition \citep{lu2015solvent, lu2016influence} on the average oil droplet size were also qualitatively investigated. The mutual interaction between a multitude of surface droplets and the resulting effect on the growth dynamics was also studied \citep{xu2017collective, dyett2018growth}. Despite all of these studies, a thorough understanding of the oversaturation pulse -- which is crucial to droplet nucleation \& growth -- is still lacking, because of its transient nature and a lack of means to directly measure it.

To have a \textit{quantitative} understanding of the oversaturation pulse, we  study the total amount of oversaturated oil inside a Hele-Shaw like microfluidic channel, by measuring the total volume of the oil droplets that nucleate out from it using confocal microscopy. In this paper, we are neither interested in the nucleation process itself nor in the droplet morphology, since they do not help to quantify the oversaturation pulse. A theoretical model for the total nucleated oil volume is developed, based on the ternary phase diagram and Taylor-Aris dispersion. The model accurately predicts the scaling behavior of the total volume $V$ of oil with respect to the P\'eclet number, $V\propto Pe^{1/2}$, including the prefactor, in which the influence of the solution composition and channel geometry is reflected. However, to compare the prefactor with the experiments, we need to prevent the oversaturated oil -- especially the bulk droplets that nucleated out of it -- from leaving the channel. To achieve this, a porous region consisting of circular pillars is put in the middle of the channel. For channels with such a porous region, the prefactor can collapse different groups of data onto one universal master curve for different channel geometries and mixtures. For channels without the porous region, the measured oil volume is smaller than the theoretical prediction because some of the oversaturated oil -- including nucleated droplets in the bulk -- leaves the channel.

\section{Experimental procedure \& methods}

\begin{figure}
\centering
\includegraphics[width=0.8\linewidth]{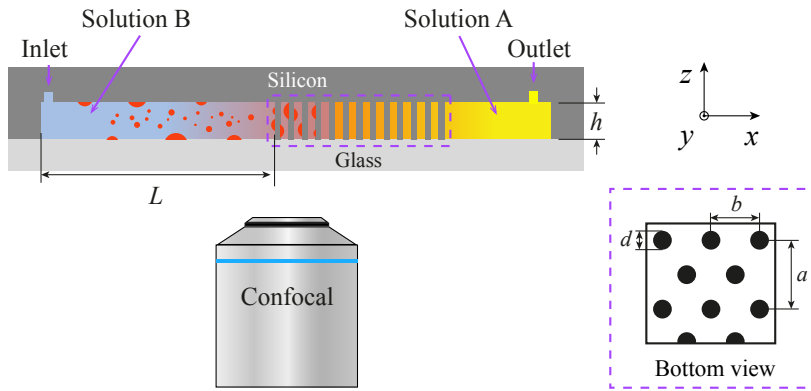}
\caption{Side view of the experimental setup. Solvent exchange is performed in a square microfluidic channel with or without the porous region, which is made of circular pillar arrays. Solution A (yellow) is decane (the oil) saturated aqueous ethanol solution and is injected into the channel first. Solution B (light blue) is decane saturated water and is injected later at a fixed pressure. In the mixture between the two solutions, oil is oversaturated so that droplets (red) nucleates out of it, both in the bulk and on the hydrophobic walls. A flowmeter is used to measure the flow rate of solution B. The channel height is $h$, width is $W$ (in $y$ direction, not shown), and the channel length before the porous region is $L$. Solvent exchange is observed from below by confocal microscopy. The inset shows the bottom view. Pillar diameter $d$ and pillar spacing $a$ and $b$ are varied to change the porosity $\phi=1-\pi d^2/(2ab)$. See table \ref{table1} for the parameters.}
\label{fig:1setup}
\end{figure}

Solvent exchange is performed in a thin square channel with various height $h$, width $W$ and length $L$, see figure \ref{fig:1setup} for the definitions and table \ref{table1} for the parameters. The microfluidic channel is made of a glass wafer covered on a silicon wafer which is decorated with an inlet, an outlet, and some of them a porous region in the center. The porous region is made of an array of circular pillars with pillar diameter $d$ ranging from \SI{6.7}{\micro\meter} to \SI{8.7}{\micro\meter} (see figure \ref{fig:1setup}, inset), and pillar spacing in the transverse and axial directions $a$ and $b$ are varied to change the porosity $\phi=1-\pi d^2/(2ab)$ (see table \ref{table1} for parameters). The length between the inlet \& the porous region is $L$. For ``smooth'' channels, i.e., those without the porous region, $L$ is the entire length of the channel. The whole channel is made hydrophobic by OTS (octadecyltrichlorosilane) coating: \SI{1}{M} hydrochloric acid is first pumped through the chip at \SI{50}{\micro\liter/\minute} for \SI{20}{min} by using a syringe pump (Harvard, PHD 2000). The chip is then put in a vacuum chamber at \SI{1.8}{\milli\bar} for overnight to dry. A solution of OTS dissolved in hexadecane (Sigma-Aldrich, $\geq\SI{99}{\%}$) at \SI{0.4}{v/v\%} is pumped through the chip at \SI{50}{\micro\liter/\minute} for \SI{20}{min}. The chip is then sequentially cleaned by chloroform, toluene, and ethanol, and finally dried in vacuum for use.

\begin{table}
\vspace{-0.2cm} 
	\begin{center}
	\begin{tabularx}{\textwidth}{XXXXXXXr}
Chip No. & $h$ (\si{\micro\meter}) & $L$ (\si{mm}) & $W$ (\si{\micro\meter})&$d$ (\si{\micro\meter}) & $a$ (\si{\micro\meter}) & $b$ (\si{\micro\meter}) & $\phi=1-\pi d^2/(2ab)$ \\ [-3.5pt]
\hline \\[-12pt]
1 & 15.5 & 4 & 1280 & 8.7 & 30 & 20 & 0.80\\
2 & 19.7 & 4 & 640   & 6.7 & 30 & 20 & 0.88\\
3 & 16.8 & 4 & 640   & 8.7 & 30 & 20 & 0.80\\
4 & 15.5 & 7 & 1280 & 8.7 & 30 & 20 & 0.80\\
5 & 19.7 & 4 & 640   & 6.7 & 30 & 20 & 0.88\\
6 & 15.5 & 13 & 1280 & --   & --   & --  & 1\\[-4pt]
\hline \\[-15pt]
\end{tabularx}
\caption{Geometrical parameters of the microfluidic chips. See figure \ref{fig:1setup} for a definition of $h$, $L$, $W$, $d$, $a$ and $b$.}
\label{table1}
\end{center}
\vspace{-0.15cm} 
\end{table}

\color{black}
Solution A, which is rich in oil, consists of decane (oil) saturated aqueous ethanol solution. To make a solution A with the desired concentration, its  ethanol-to-water weight ratio $w_\mathrm{e,A}/w_\mathrm{w,A}$ is first determined. The solution is prepared by first mix $\sim\SI{400}{g}$ mixture of ethanol (Sigma-Aldrich, $\geq\SI{99.8}{\%}$) and water (Milli-Q) with this specific weight ratio $w_\mathrm{e,A}/w_\mathrm{w,A}$, then add in decane (Sigma-Aldrich, $\geq\SI{95}{\%}$) until phase separation is observed, so that the solution is saturated. The weight fractions of each species are calculated from the actual ethanol-to-water weight ratio $w_\mathrm{e,A}/w_\mathrm{w,A}$ in the solution. The oil weight fraction $w_\mathrm{o,A}$ of solution A is increased by increasing its ethanol-to-water weight ratio $w_\mathrm{e,A}/w_\mathrm{w,A}$, since the solubility of oil (decane) in ethanol is higher than in water. Finally, solution A is labelled yellow by adding a small amount of perylene (Sigma-Aldrich, $\geq\SI{99}{\%}$) at \SI{0.2}{mg/mL}.

Solution B, which is poor in oil, is made of decane saturated water and is labelled light blue by Rhodamine 6G (Sigma-Aldrich, 99\%) at \SI{0.2}{mg/mL}. Its oil weight fraction is $w_\mathrm{o,B}=\SI{5.2e-8}{}$ at \SI{25}{\degreeCelsius}.

Solution A is first injected to fill the entire channel, then solution B is injected at a constant driving pressure to perform the solvent exchange: An oil oversaturation pulse is generated in the mixture of solutions A and B, which leads to oil droplets nucleation both in the bulk and on the hydrophobic surfaces in the channel. The contact angle of oil in water is $\theta=15\pm3\si{\degree}$ on the same treated silicon surface and $\theta=11\pm2\si{\degree}$ on the same treated glass (see SI for more details). The flow rate $Q$ of solution B, measured by a flowmeter (ML120V21, Bronkhorst, Netherlands), is varied by changing the driving pressure. The P\'eclet number of the flow is calculated by $Pe_h=\overline u h/D$, where $\overline u=Q/(Wh)$ is the average velocity of solution B, and a typical diffusivity of water in ethanol $D=\SI{0.84e-9}{m^2/s}$ is used. 

After about \SI{0.3}{mL} (more than 1000 times of the channel volume $\approx \SI{0.25}{\micro\liter}$) of solution B is injected, the injection is stopped by closing the valve, and a 3D scan of the whole channel is recorded by confocal microscopy (Nikon A1, Nikon, Japan) from below. 

\section{Experimental results}

\begin{figure}
\centering
\includegraphics[width=0.7\linewidth]{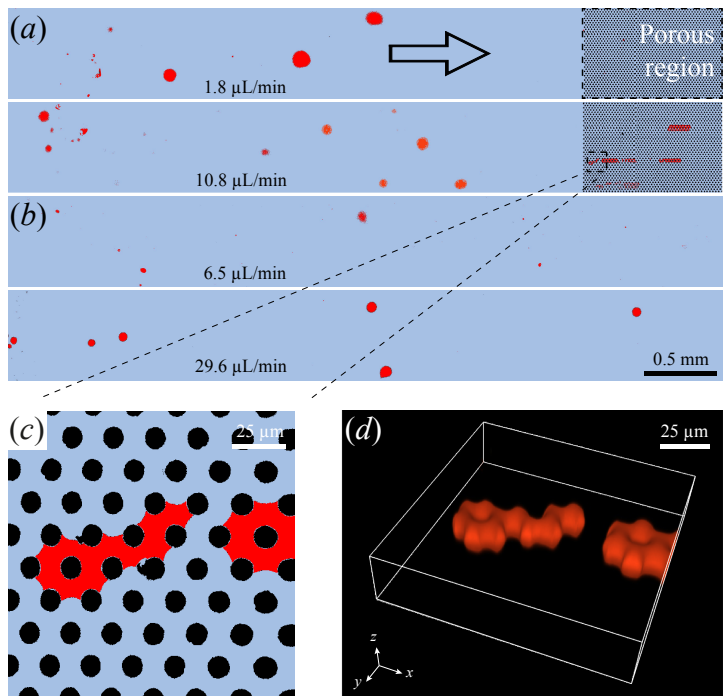}
\caption{($a$) Typical mid-plane snapshots of the upstream part of Chip No. 1, see table \ref{table1} for the geometrical parameters. Red signals oil, light blue signals water, and black the pillars. The flow rates are \SI{1.8}{\micro\liter/min} and \SI{10.8}{\micro\liter/min}. The arrow indicates the direction of the flow. ($b$) Typical mid-plane snapshots of the upstream part of Chip No. 6. The flow rates are \SI{6.5}{\micro\liter/min} and \SI{29.6}{\micro\liter/min}, respectively. The scale bar indicates \SI{0.5}{mm}. Oil droplets are found both before the porous region and inside the porous region. ($c$) A closer look at the porous region of Chip No. 1, where the circular pillars are clearly seen. ($d$) 3D confocal image of the morphology of oil in the same area, resolution is $0.28\times 0.28 \times \SI{0.85}{\micro\cubic\meter\per pixel}$ in the $x$, $y$, and $z$ direction, respectively.}
\label{fig:2Snapshot}
\end{figure}

Typical mid-plane snapshots of the upstream part of the channel with the porous region (Chip No. 1, see table \ref{table1} for the geometrical parameters) are shown in figure \ref{fig:2Snapshot}($a$). The channel without porous region (Chip No. 6, see table \ref{table1}) is shown in figure \ref{fig:2Snapshot}($b$). Black signals pillars, light blue signals water, and red signals oil (because ethanol must have been dissolved in and washed away by the excess amount of water). In general, more oil droplets (red) are found in the channel after the solvent exchange. Furthermore, oil droplets are observed before and inside the porous region (black), but not behind the porous region. However, for channels without the porous region, droplets are observed in the entire channel (see SI for the snapshots of the full channel). A closer look at the porous region is shown in figure \ref{fig:2Snapshot}($c$), where the oil droplets and the circular pillars are clearly shown. A typical 3D scan of the same area is shown in figure \ref{fig:2Snapshot}($d$). Note that in this work, we only focus on the total oil volume $V$, but not on the droplet morphology.

Solvent exchange is performed for all the 6 chips (see table \ref{table1} for the geometrical parameters) at different $w_\mathrm{e,A}$ and flow rate $Q$. After the solvent exchange, 3D scans of the whole channel, similar to that shown in figure \ref{fig:2Snapshot}($d$) are performed. The total volume $V$ of these droplets is measured by counting the number of red pixels of the 3D confocal image and then multiplied by the volume of one pixel, then they are plotted against $Pe_h$ in figure \ref{fig:5}($a$)\&($b$) in log-log scale. Results for channels with the porous region (chip No. 1-5) are shown in figure \ref{fig:5}($a$), and results for channels without the porous region (chip No. 6) are shown in figure \ref{fig:5}($b$). It is found that for all the chips, $V$ increases with the P\'eclet number as $\propto Pe_h^\alpha$, with $\alpha\approx 0.50$, see table \ref{table2} for details. $V$ also increases with $w_\mathrm{o,A}$, i.e., the more oil is in solution A, the more oil is nucleated. In the next subsection we will develop a theoretical model to quantitatively account for these two observations.

\begin{table}
\vspace{-0.2cm}
\begin{center}
\begin{tabularx}{\textwidth}{XXXlr}

Chip No. &$w_\text{e,A}$ &$w_\text{o,A}$& exponent $\alpha$ & \SI{95}{\%} confidence bounds  \\[-3.5pt]
\hline \\[-12pt]
1& 0.821 &	0.063 & 0.49 & $\pm\,0.05$ \\ 
2& 0.821 &	0.063 & 0.49 & $\pm\,0.04$ \\
3& 0.821 &        0.063&  0.50 & $\pm\, 0.04$ \\
4& 0.754 &	0.017 & 0.49 & $\pm\,0.05$ \\ 
5& 0.754 &        0.017 & 0.51 & $\pm\,0.04$ \\ 
6& 0.821 & 0.063	 & 0.51 & $\pm\,0.05$ \\
6& 0.754 & 0.017	 & 0.51 & $\pm\,0.01$ \\[-4pt]

\hline \\[-18pt]
\end{tabularx}
\caption{Control parameters of the solvent exchange, and the best fitted power laws: $V\sim Pe_h^\alpha$. $w_\text{o,A}$ is the oil weight fraction of the good solvent, $w_\text{e,A}$ is the corresponding ethanol weight fraction, and $\alpha$ is the fitted exponent of the power law. \SI{95}{\%} confidence bounds of the fittings are also shown. The average exponent $\alpha$ is $0.50\pm0.01$.}
\label{table2}
\end{center}
\vspace{-0.15cm}
\end{table}

\section{Theoretical model}

\begin{figure}
\centering
\includegraphics[width=0.96\linewidth]{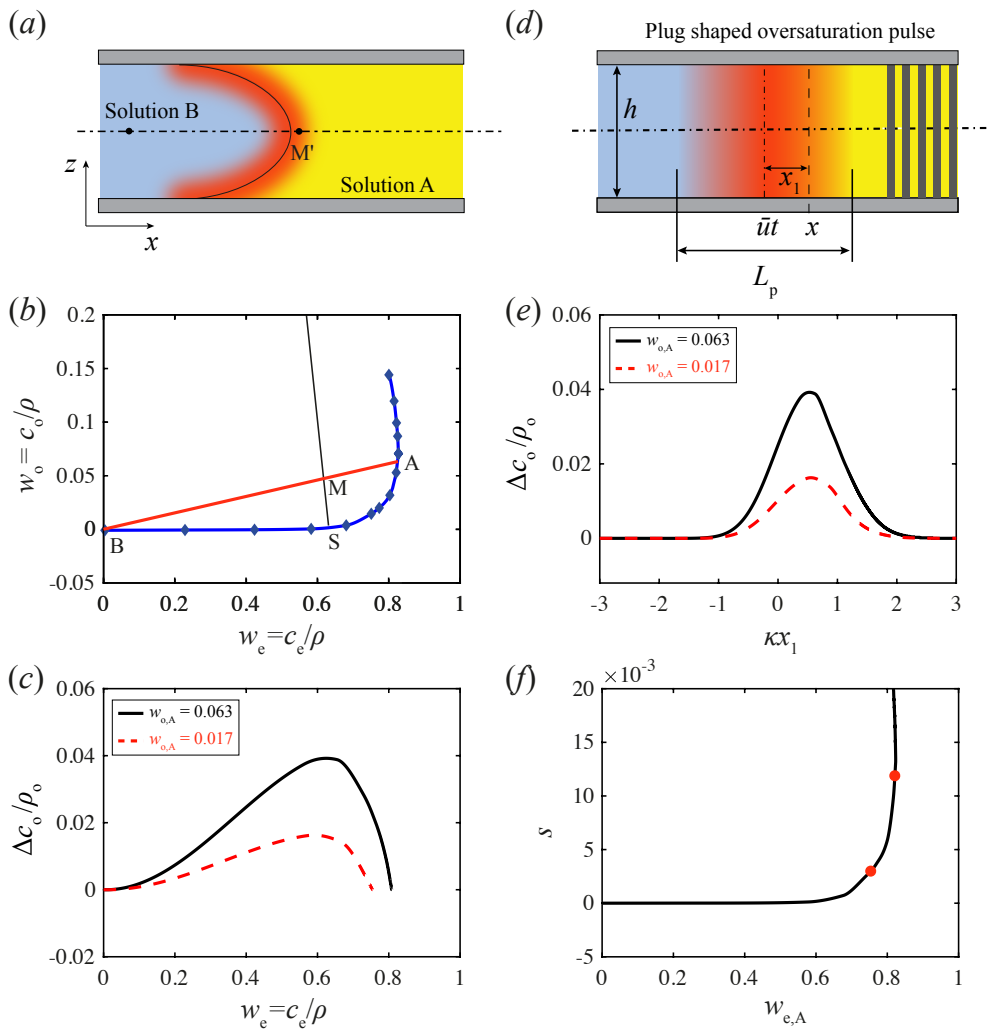}
\caption{($a$) Sketch of the oversaturation pulse (red region) during solvent exchange. Solution A is on the right hand side, being displaced by solution B. Liquid parcel M$^\prime$ is on the oversaturation pulse. The concentrations of solutions A, B, and liquid parcel M$^\prime$ are represented by A, B and M in ($b$), respectively. ($b$) Phase diagram of the decane-ethanol-water ternary system. The blue curve is the binodal curve (saturation curve of oil) fitted from the data points (blue diamond) measured as described in \cite{tan2016evaporation}. Red line AB is the diffusion path, black line MS is the dilution curve, which is a straight line passing through coordinate $(0,1)$. ($c$) Oil oversaturation normalized by the oil density, $\Delta c_\mathrm{o}/\rho_\mathrm{o}$, as a function of the normalized ethanol concentration $w_\mathrm{e}=c_\mathrm{e}/\rho$, at different initial conditions: Black solid line for $w_\mathrm{o,A}=0.063$, and red dashed line for $w_\mathrm{o,A}=0.017$. ($d$) Plug shaped oversaturation pulse because of Taylor-Aris dispersion which happens in a long \& thin channel. $x_1=x-\overline{u}t$ is the distance to the center of the plug. $\overline{u}$ is the average flow velocity, and $L_\mathrm{p}$ is the length of the plug. The porous region is represented by the vertical black pillars. ($e$) The normalized oil oversaturation $\Delta c_\mathrm{o}/\rho_\text{o}$ as a function of normalized length $\kappa x_1$, for $w_\mathrm{o,A}=0.063$ (black solid line), and $w_\mathrm{o,A}=0.017$ (red dashed line), respectively. The normalized length $\kappa x_1$ is a function of time $t$ and space $x$. ($f$) The calculated prefactor $s$ as a function of $w_\mathrm{e,A}$. The two different initial conditions with $w_\mathrm{o,A}=0.063$ and 0.017 are indicated by the red dots.}
\label{fig:4model}
\end{figure}

Figure \ref{fig:4model}($a$) shows the schematic of the initial oversaturation pulse (red shaded region), which is a mixture of the two solutions due to advection and diffusion. The initial oversaturation pulse takes a parabolic shape in laminar flow, which is the case here. It broadens because of (Taylor-Aris) diffusion of the three components: oil, ethanol and water. The concentration of the mixture changes continuously from that of the solution A to that of solution B \citep{ruschak1972spontaneous}. Figure \ref{fig:4model}($b$) is the phase diagram of the oil-ethanol-water system, it shows the concentrations of the solutions \& the mixture, with the ethanol weight fraction $w_\mathrm{e}$ and the oil weight fraction $w_\mathrm{o}$ being the $x$ and $y$ axes, respectively. Then the water weight fraction of the solution/mixture can be calculated by $w_\mathrm{w}=1-w_\mathrm{e}-w_\mathrm{o}$. The blue curve shows the concentrations of (oil) saturated solutions, this is the so called ``binodal curve''. The curve is fitted from the measured data points (see Appendix \ref{appPCHIP} for details). Then, the concentrations of solutions A and B are on the binodal curve, denoted by points A and B in figure \ref{fig:4model}($b$). The concentrations of the mixture lie on a curve connecting points A and B, denoted by the red curve AB -- this is the so called ``diffusion path'' \citep{ruschak1972spontaneous}. All the possible concentrations of the liquid -- both of the solutions and the mixture -- lie on this curve. The oil concentration $c_\mathrm{o}$ in the mixture is higher than its saturated oil concentration $c_\mathrm{o,s}$, thus the mixture is oversaturated with oil. The (absolute) oil oversaturation is then denoted as 
\begin{equation}
\Delta c_\mathrm{o}\equiv c_\mathrm{o}-c_\mathrm{o,s}.
\label{eq:Xi}
\end{equation}

We now use mass-per-volume concentrations $c$ in the equations for simplicity but keep using weight fractions $w$ elsewhere. Notice that the weight fractions $w$ are the concentrations $c$ nomalized by the density of the liquid $\rho$: $w=c/\rho$.

For any oversaturated liquid parcel M$^\prime$ in the mixture as represented by point M on the diffusion path, when the oversaturated oil nucleates, it is considered that only oil nucleates out of the mixture, and the ethanol-to-water ratio in the mixture is kept constant \citep{lu2016influence}. The concentration of the mixture moves along the so-called ``dilution curve'' \citep{lu2016influence} to point S on the binodal curve. The dilution curve MS is a straight line passing through point $(0,1)$ in figure \ref{fig:4model}($b$) (this point means pure oil in the phase diagram, see SI \& Appendix for more details), as shown by the black solid line. Therefore, the (absolute) oil oversaturation of this liquid parcel M$^\prime$ is:
\begin{equation}
\Delta c_\mathrm{o}(\mathrm{M})=(c_\mathrm{o}-c_\mathrm{o,s})|_\mathrm{M}=c_\mathrm{o}(\mathrm{M})-c_\mathrm{o}(\mathrm{S}).
\label{eq:OverSatM}
\end{equation}

The exact shape of the diffusion path is determined by the diffusion speeds of the three components. In a simplified case, we assume that the diffusion of each component only depends on its own concentration gradient. To get a first order approximation of the problem, we further assume that the diffusion coefficients of oil and water are equal: $D_\text{o}=D_\text{w}\equiv D$ . Then the diffusion path AB becomes a straight line (see \cite{ruschak1972spontaneous} and Appendix \ref{appTransform}. These two assumptions are reasonable since the oil concentration is small in all the tested cases: $w_\mathrm{o,A}\leq0.063$, and the errors induced by them are small, see Appendix \ref{appDiff}), and $c_\mathrm{o}(\mathrm{M})$ becomes linearly dependent on the ethanol concentration $c_\mathrm{e}$ of the arbitrary liquid parcel M$^\prime$. Since point M is on the diffusion path AB, we have $0\leq c_\mathrm{e}\leq c_\mathrm{e,A}$, where the initial condition $c_\mathrm{e,A}$ is the ethanol concentration of solution A. $c_\mathrm{o}(\mathrm{S})$ can also be expressed as a function of $c_\mathrm{e}(\mathrm{M})$ by finding the intersection of the dilution curve (passing through point M) and the binodal curve. Then the (absolute) oil oversaturation $\Delta c_\mathrm{o}$ of any liquid parcel in the channel can be expressed as a function of its ethanol concentration $c_\mathrm{e}$:
\begin{equation}
\Delta c_\mathrm{o}=\Delta c_\mathrm{o}(c_\mathrm{e}), \quad 0\leq c_\mathrm{e}\leq c_\mathrm{e,A}.
\label{eq:XionCe}
\end{equation}

This function of course also depends on the initial condition $c_\mathrm{e,A}$. Eq.(\ref{eq:XionCe}) is calculated numerically (see SI for more details), and the oil oversaturation normalized by the oil density, $\Delta c_\mathrm{o}/\rho_\mathrm{o}$, is plotted as a function of the normalized ethanol concentration $w_\mathrm{e}=c_\mathrm{e}/\rho$ in figure \ref{fig:4model}($c$), for different initial conditions $w_\mathrm{o,A}=c_\mathrm{o,A}/\rho$: 0.063 and 0.017, respectively. Here $\rho$ is the density of the liquid parcel (see SI for the calculation of $\rho$), and $\rho_\mathrm{o}=\SI{730}{kg/m^3}$ is the density of the oil (decane). 

The oversaturation pulse shown in figure \ref{fig:4model}($a$) evolves as a function of space $\boldsymbol{x}$ and time $t$, so that both the oil oversaturation $\Delta c_\mathrm{o}$ and $c_\mathrm{e}$ also depend on space and time: $\Delta c_\mathrm{o}[c_\mathrm{e}(\boldsymbol{x},t,c_\mathrm{e,A})]$. As mentioned previously, the total oil volume $V$ is nucleated from \textit{all} the oil oversaturation generated in the liquid. The total oil volume $V$ in the porous media can then be calculated by integrating the oil oversaturation over the entire channel volume, at the time just before the oversaturation pulse reaches the porous region:
\begin{equation}
V(t,c_\mathrm{e,A})= \displaystyle \frac{1}{\rho_\mathrm{o}}\int_{\Omega}\,  \Delta c_\mathrm{o}[c_\mathrm{e}(\boldsymbol x,t,c_\mathrm{e,A})] \,\mathrm{d}\boldsymbol x.
\label{eq:VtotGeneral}
\end{equation}

The spatial and temporal distribution of ethanol concentration $c_\mathrm{e}(\boldsymbol x,t,c_\mathrm{e,A})$ of the parabolic shaped oversaturation pulse is non-trivial. However, the channel used here is long and thin. The time scale for the axial advection is $\tau_\mathrm{a} \sim L/\overline u \geq \SI{4}{s}$, which is much larger than the time scale of the vertical diffusion $\tau_\mathrm{v} \sim (h/2)^2/D=\SI{0.12}{s}$. In this situation, the concentration gradient in the vertical direction is smoothed out because of Taylor-Aris dispersion \citep{taylor1953dispersion, aris1956dispersion, aris1959dispersion}, and the oversaturation pulse becomes like a plug, as shown in figure \ref{fig:4model}($d$). 

The analytical solution of the ethanol concentration $c_\mathrm{e}(\boldsymbol x, t,c_\mathrm{e,A})$ for pure water entering a thin square channel which at time $t=0$ contains only an aqueous ethanol solution with (uniform) concentration $c_\text{e,A}$ (no oil present) in this plugged regime is \citep{taylor1953dispersion}:
\begin{equation}
c_\text{e}(\boldsymbol x,t,c_\text{e,A})=\frac{1}{2}[1+\text{erf}(\kappa x_1)]\cdot c_\text{e,A},
\label{eq:Ce}
\end{equation}
where $x_1=x-\overline{u}t$ is the distance to the central position of the plug, as shown in figure 2($d$). $\kappa = \frac{1}{2} k^{-\frac{1}{2}}t^{-\frac{1}{2}}$ is the wavenumber (with unit $\si{m^{-1}}$), where $k=h^2\overline u^2/(210D)$ for a thin square channel of thickness $h$ \citep{dorfman2001comment}. The wavenumber $\kappa$ has the temporal dependence and its inverse scales as the length of the plug: $L_\mathrm{p}\sim 1/\kappa$. 

For the solvent exchange in our case, oil is present in all the liquid, but Eq.(\ref{eq:Ce}) still holds here because of the previous assumption $D_\text{o}=D_\text{w}\equiv D$. With this assumption, any portion of the water can be replaced by oil without influencing the above equation. Substituting Eq.(\ref{eq:Ce}) into Eq(\ref{eq:XionCe}), we have:
\begin{equation}
\Delta c_\mathrm{o}=\Delta c_\mathrm{o}(\kappa x_1, c_\mathrm{e,A}).
\label{eq:Xionkappa}
\end{equation}

Eq.(\ref{eq:Xionkappa}) is solved numerically, and figure \ref{fig:4model}($e$) shows how the normalized oil oversaturation $\Delta c_\mathrm{o}/\rho_\mathrm{o}$ changes as a function of $\kappa x_1$ at the three different initial conditions. It is worth noting that the oil oversaturation is not symmetric, but its peak position is to the right ($\kappa x_1\approx 0.55$), or in other words, more to the downstream. This asymmetry originates from the asymmetry in $\frac{\Delta c_\mathrm{o}}{\rho_\mathrm{o}}(w_\mathrm{e})$ shown in figure \ref{fig:4model}($c$), whose asymmetry originates from the shape of the binodal curve shown in figure \ref{fig:4model}($b$). Also, the oil oversaturation is higher everywhere in the channel for the case with higher initial oil concentration (larger $w_\mathrm{o,A}$). 

Since the oversaturation pulse becomes a plug-like shape, which means the vertical concentration gradient is negligible,  Eq.(\ref{eq:VtotGeneral}) can be further simplified as the integration along the single dimension $x$. Substituting Eq.(\ref{eq:Xionkappa}), we have:

\begin{equation}
V(t,c_\mathrm{e,A})= \displaystyle \frac{1}{\rho_\text{o}}Wh\int_{-\infty}^{\infty}\,  \Delta c_\text{o}(\kappa x_1, c_\mathrm{e,A}) \,\text{d}x.
\label{eq:1D}
\end{equation}

Substitute $t=L/\overline u$ into $\kappa$ as the time when the oversaturation pulse is just entering the porous region. Then the integration is performed with respect to $\kappa x_1$, which leads to:
\begin{equation}
V(c_\mathrm{e,A})=s(c_\mathrm{e,A})\cdot W\displaystyle h^{\frac{3}{2}}\displaystyle L^{\frac{1}{2}}\cdot Pe_h^{\frac{1}{2}},
\label{eq:VFinal}
\end{equation}
where 
\begin{equation}
s(c_\mathrm{e,A})=\displaystyle\frac{2}{\sqrt{210}}\frac{1}{\rho_\mathrm{o}}\int_{-\infty}^{\infty}  \Delta c_\mathrm{o}(\kappa x_1, c_\mathrm{e,A})\,\text{d}(\kappa x_1)
\label{eq:s}
\end{equation}
is a dimensionless prefactor which is proportional to the area under the curve $\Delta c_\mathrm{o}(\kappa x_1, c_\mathrm{e,A})$, as shown in figure \ref{fig:4model}($e$). 

From eq.(\ref{eq:VFinal}) we see that the power law dependence of $V$ on $Pe_h$ is indeed predicted to be $1/2$. The prefactor $s$ is shown in figure \ref{fig:4model}($f$) as a function of normalized initial condition $w_\mathrm{e,A}=c_\mathrm{e,A}/\rho$. The two red dots correspond to the initial conditions $w_\mathrm{o,A}=0.063$ and 0.017, respectively. It is worth noting that $s$ changes sharply within this range.

Eq.(\ref{eq:VFinal}) with eq.(\ref{eq:s}) is the main theoretical result of our paper. It is universal in the sense that it includes all the factors that could influence the total amount of nucleated oil: the flow rate, the channel geometry, and the solution composition. 

Note that in the model, only Taylor dispersion in the post-free region is considered, because the porous structure immediately interrupts further development of the oversaturation pulse. Further dispersion/mixing in the porous region is negligible: The length of the oversaturation pulse, which covers \SI{99}{\%} of the oil oversaturation (see figure \ref{fig:4model}($e$)), is calculated to be $L_\mathrm{p}=3.56/\kappa=0.49\sqrt{Lh}\cdot\sqrt{Pe_h}>\SI{0.44}{mm}$, much larger than the pore size $\approx\SI{20}{\micro\meter}$. That means that $c_\mathrm{e}$ does not really change on the length scale of the pore size, thus the porous media does not induce further dispersion/mixing. On the other hand, though in the theoretical model the length of the post-free region $L$ is used to calculate the total oil volume, this does not mean that the oversaturation would accumulate in the pulse and then all of a sudden nucleate into droplets at some time point, for example, at $t=L/\overline{u}$ when the center of the oversaturation pulse is at the entrance of the porous region. Instead, as the oversaturation pulse develops while moving downstream, the (newly generated) oversaturation nucleates into droplets along the way, so that some of the droplets are observed upstream of the porous region.

\section{Comparison between universal theoretical result \& experiments}

\begin{figure}
\centering
\includegraphics[width=0.9\linewidth]{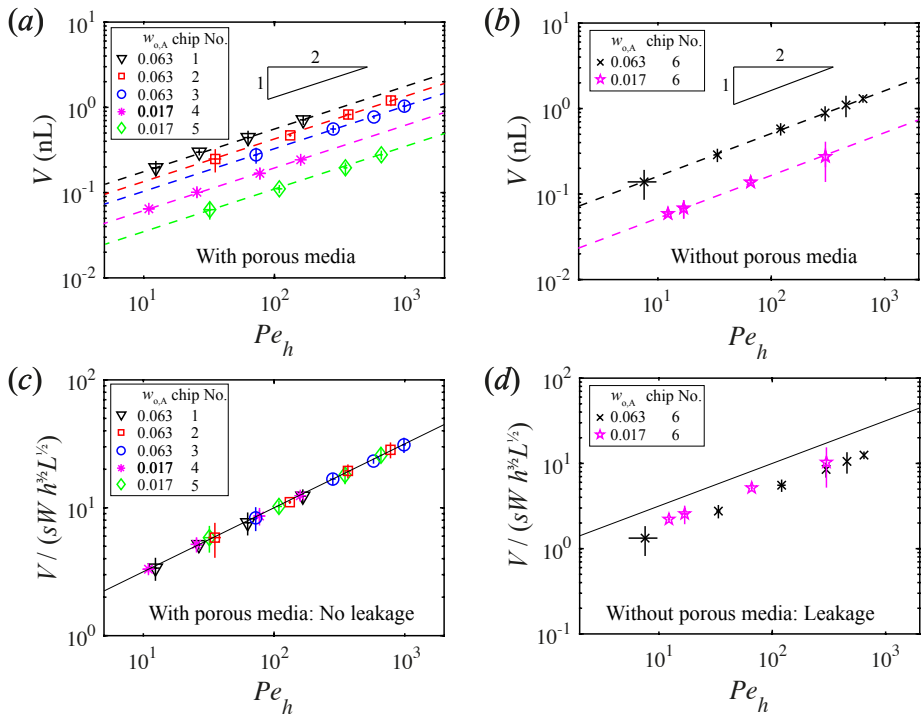}
\caption{($a$) \& ($b$): Measured total volume of oil in the entire channel $V$ as functions of the P\'eclet number $Pe_h=h\overline u/D$, for different initial conditions -- oil weight fraction of solution A represented by $w_\mathrm{o,A}$, and different flow rates as represent by $Pe_h$, in different chips. Results for the channels with the porous region (chip No. 1-5) are shown in ($a$), and results for the channel without the porous region (chip No. 6) are shown in ($b$). For all the results, the total oil volume increases as $V\sim Pe_h^{\alpha}$, with $\alpha\approx0.50$. Dashed lines are the best fit for each group. See table \ref{table2} for the solution compositions of the solvent exchange and the best fitted power laws. ($c$) \& ($d$): $V$ non-dimensionalized by \mbox{$s\cdot Wh^{3/2}L^{1/2}$} and plotted as functions of $Pe_h$. Results for the channels with the porous region (chip No. 1-5) are shown in ($c$), and results for the channel without the porous region (chip No. 6) in ($d$). The black solid lines are calculated from Eq.(\ref{eq:VFinal}). Each data point is the average of five measurements and error bars are the standard deviation.}
\label{fig:5}
\end{figure}

To compare the universal theoretical result with the experimental results, the oil volume $V$ measured for various chips and mixtures shown in figure \ref{fig:5}($a$)\&($b$) are normalized by $s\cdot Wh^{3/2}L^{1/2}$, and then plotted against $Pe_h$ in figure \ref{fig:5}($c$)\&($d$) in log-log scale. Results for channels with the porous region (chip No. 1-5) are shown in figure \ref{fig:5}($c$), and results for channels without the porous region (chip No. 6) are shown in figure \ref{fig:5}($d$). It is found that for chips with the porous region, the calculated prefactors ($s=\SI{1.2e-2}{}$ and \SI{3e-3}{})  can indeed collapse \textit{all} data onto one master curve, regardless of the channel geometry and composition of the mixture. This is in agreement with the calculated prefactor of our theoretical model. However, for channels without the porous region, the measured oil volume $V$ is smaller than the theoretical prediction -- which is not true for channels with the porous region. This confirms that the porous region can indeed prevent the oversaturated oil from leaving the channel, by which the comparison between experiment \& theory is enabled.

\section{Conclusions and outlook}

We have performed solvent exchanges in thin square channels with and without a porous region in the middle, at different initial conditions $w_\mathrm{e,A}$, flow rates $Q$ and channel geometries. The total volume of these oil droplets $V$ is measured by confocal microscopy, and is found to increase with P\'eclet number as $\propto Pe_h^{\alpha}$, with $\alpha\approx 0.50$. A theoretical model is then developed, based on the ternary phase diagram and Taylor-Aris dispersion, to predict the total amount of oil $V$ nucleated from solvent exchange. The theory indeed predicts a power law dependence $V\propto Pe_h^{1/2}$. In addition, the influence of the channel geometry and the initial condition $w_\mathrm{o,A}$ can also be calculated and included in the model to give a complete prediction of the oil volume $V$ on one \textit{universal} curve, thanks to the porous region which can prevent the oversaturated oil from leaving the channel. This model is found to be able to predict the total nucleated oil volume $V$, irrespectively of the channel geometry \& initial mixture.

The findings of this work contribute to a better understanding of the solvent exchange, and could guide further design and research on this topic. First, a porous media may serve as a good tool to collect all the oversaturation, providing a macroscopic approach to study this bottom-up method. 

The results of this paper encourage us to propose a three-step approach to study solvent exchange: (i) Find the concentration distribution by solving the advection-diffusion equations. The boundary conditions are defined by the flow geometries and the initial conditions are set by the initial solution concentrations. (ii) Based on the concentration distribution, the oversaturation distribution can be calculated by applying the knowledge of the phase diagram. (iii) Investigate the quantity that is of interest, such as the volume of the nucleated phase, the dynamical interaction between the nucleated phase and the oversaturation, etc. 

This research can be considered as a demonstration of the above proposed approach, with the total nucleated oil volume $V$ being the subject of interest, and the aid of Taylor-Aris dispersion in a long \& thin channel to obtain the analytical solution of the concentration distribution. For more general cases where the oversaturation pulse is not a plug, analytical solutions of the concentration distribution may be difficult to get, then numerical simulations should be employed to finish the first step. It is also easier to incorporate the ternary (or multicomponent) diffusion effect in numerical simulations. Moreover, the dynamical interaction between the nucleation and the oversaturation pulse can also be incorporated in numerical simulations.

\begin{acknowledgements}
\section*{Acknowledgements}
We thank Xuehua Zhang and Chao Sun for valuable discussions, and Hai Le The for the SEM Images. We thank the Micro-Nano Fabrication Laboratory of Peking University for providing the chips. We acknowledge support from the Netherlands Center for Multiscale Catalytic Energy Conversion (MCEC), an NWO Gravitation programme funded by the Ministry of Education, Culture and Science of the government of the Netherlands, and D.L.'s ERC-Advanced Grant under project number 740479.
\end{acknowledgements}

\section*{Declaration of Interests}
The authors report no conflict of interest.

\appendix

\section{Fitted binodal from the data points}\label{appPCHIP}
The X- and Y-coordinate in Fig.\ref{fig:4model}($b$) are actually $w_\mathrm{e}=c_\mathrm{e}/\rho$ and $w_\mathrm{o}=c_\mathrm{o}/\rho$. The binodal $w_\text{o,s}(w_\text{e})$ is fitted by two Piecewise Cubic Hermite Interpolating Polynomial (PCHIP) through the data points, which are measured by titration following the procedure as described by \cite{tan2016evaporation}. 

For the first part (first 10 points, lower half), the polynomial is:
\begin{equation}
w_\text{o,s}=w_\text{o,s}(w_\text{e})=\sum_{i=1}^{10}P_i(w_\text{e})
\label{eq:Cs}
\end{equation}
where
\begin{equation}
P_i(w)=a_i(w-w_i)^3+b_i(w-w_i)^2+c_i(w-w_i)+d_i
\label{eq:Pi}
\end{equation}
is valid on the 9 intervals between the 10 data points. Here $w_i$ is the ethanol weight fraction of the $i^\text{th}$ data point. 
For the second part (last 5 points, upper half), the polynomial is:
\begin{equation}
w_\text{o,s}=w_\text{o,s}(w_\text{e})=\sum_{i=11}^{15}P_i(w_\text{e})
\label{eq:Cs}
\end{equation}
where $P_i(w)$ has the same form as Eq.\ref{eq:Pi}.
Parameters are shown in table \ref{table3}.

\begin{table}
\vspace{-0.2cm}
\begin{tabularx}{\textwidth}{ccXXXXr}

$i$ &$\quad$& $w_i$ & $a_i$ & $b_i$ & $c_i$ & $d_i$ \\[-3.5pt] 
 \hline \\[-12pt]

1 &$\quad$& 0          & \SI{2.0712e-4}{} & $\!\!\!\!\!-\SI{4.6769e-4}{}$ & $\SI{1.1e-3}{}$ & $\SI{9e-9}{}$ \\ 
2 &$\quad$& 0.2248 &  $0.0187$ & $\!\!\!\!\!-0.0032$   & $0.0012$ & $\SI{2.6e-4}{}$ \\ 
3 &$\quad$& 0.4198 & $0.0562$ &  $0.0085$   & \SI{2.1e-3}{} & $\SI{5.2e-4}{}$ \\ 
4 &$\quad$& 0.5787 & $\!\!\!\!\!-0.0937$ & $0.2582$ & $0.0091$ & $0.0013$ \\ 
5 &$\quad$& 0.6770 & $\!\!\!\!\!\!\!\!-10.9639$ & $2.1452$ & $0.0571$ & $0.0046$ \\ 
6 &$\quad$& 0.7473 & $\!\!\!15.1584$ & $1.7834$ & $0.1962$ & $0.0154$ \\ 
7 &$\quad$& 0.7696 & $\!\!\!\!\!\!140.7443$ &  $\!\!\!\!\!-0.8690$ & $0.2987$ & $0.0209$\\
8 &$\quad$& 0.7994 & $\!\!\!\!\!\!121.2367$ & $\!\!\!29.3490$ & $0.6207$ & $0.0327$\\
9 &$\quad$& 0.8172 & $\!\!\!\!-\SI{1.5503e4}{}$ & $\!\!\!\!\!\!262.6259$ & $1.7831$ & $0.0538$\\
10&$\quad$&0.8235 & & & &\\
11&$\quad$&0.7968 & $\SI{1.6904e3}{}$ & $\!\!\!\!\!\!\!-87.4670$ & $\!\!\!\!\!-0.7654$ & $0.1450$\\
12&$\quad$&0.8113 & $\SI{7.9352e3}{}$ & $\!\!\!\!\!\!\!\!\!\!-169.5336$ & $\!\!\!\!\!-2.2375$ & $0.1205$\\
13&$\quad$&0.8181 & $\!\!\!\!\!-\SI{8.0645e3}{}$ & $\!\!\!\!\!126.1344$ & $\!\!\!\!\!-3.4412$ & $0.1002$\\
14&$\quad$&0.8213 & $\SI{1.3950e5}{}$ & $\!\!\!\!-\SI{1.2406e3}{}$ & $\!\!\!\!\!-5.1258$ & $0.0877$\\
15&$\quad$&0.8235 & & & &\\[-4pt]
\hline \\[-15pt]

\end{tabularx}
\caption{Parameters of the PCHIP spline fitting of the first 8 points of the binodal. $i$ is the number of the data points. $w_i$ is the ethanol weight fraction of the $i^\text{th}$ data point. $a_i$, $b_i$, $c_i$, and $d_i$ are the coefficients.} 
\label{table3}
\vspace{-0.15cm}
\end{table}

\section{Transformation from a ternary phase diagram to the phase diagram in a cartesian coordinate}\label{appTransform}

\begin{figure}
\centering
\includegraphics[width=1\textwidth]{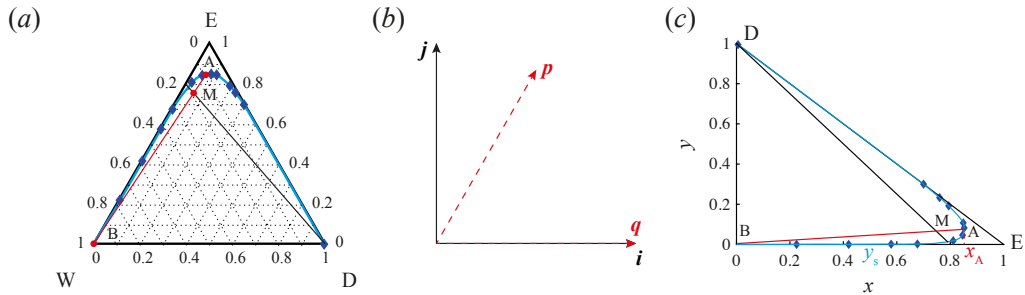}
\caption{Transformation from the ternary phase diagram to a right-triangle phase diagram in a cartesian coordinate. ($a$) The ternary phase diagram of the decane-ethanol-water system. The three apexes are E for ethanol, W for water, and D for decane. Blue curve is the binodal curve fitted from the data points (blue diamonds) taken from \cite{skrzecz1999iupac}. Red line AB is the diffusion path when $w_\mathrm{e,A}=0.837$, and black line DM is the dilution curve. ($b$) Basis ($\boldsymbol{p}$, $\boldsymbol{q}$) of the ternary phase diagram shown in red dashed arrows and basis ($\boldsymbol{i}$, $\boldsymbol{j}$) of the cartesian coordinate shown in black solid arrows. ($c$) Right-triangle phase diagram of the same system. Diffusion path AB and dilution curve DM are still straight lines. $w_\mathrm{e}$ is replaced by $x$ and $w_\mathrm{o}$ is replaced by $y$.} 
\label{fig:Trans}
\end{figure}

Figure \ref{fig:Trans}($a$) shows a typical ternary phase diagram, using the decane-ethanol-water system as an example. The three vertices E, W, D stand for ethanol, water and decane, respectively. Figure \ref{fig:Trans}($c$) shows the same phase diagram in a cartesian coordinate, with the ethanol weight fraction $w_\mathrm{e}$ being the X-coordinate and the oil weight fraction $w_\mathrm{o}$ being the Y-coordinate. Here we prove that any straight lines in the ternary phase diagram shown in figure \ref{fig:Trans}($a$) are still straight lines in figure \ref{fig:Trans}($c$). 

To prove this, we need to find the transformation matrix between these two vector spaces. Let $\boldsymbol{p}$ and $\boldsymbol{q}$ be the basis vectors of the vector space that the ternary phase diagram is in. $\boldsymbol{p}$ is parallel to line WE in figure \ref{fig:Trans}($a$), and $\boldsymbol{q}$ is parallel to line WD. Let $\boldsymbol{i}$ and $\boldsymbol{j}$ be the basis vectors of the cartesian coordinate. These two basis are put together in figure \ref{fig:Trans}($b$), with ($\boldsymbol{p}$, $\boldsymbol{q}$) shown in red dashed arrows and ($\boldsymbol{i}$, $\boldsymbol{j}$) shown in black solid arrows. Then we have:

\begin{equation}
\setlength{\arraycolsep}{4pt}
\renewcommand{\arraystretch}{1.8}
(\boldsymbol{p},\boldsymbol{q}) = (\boldsymbol{i},\boldsymbol{j})\left[
\begin{array}{cc}
   \displaystyle \frac{1}{2} & \displaystyle\frac{\sqrt{3}}{2} \\
   1 & 0 \\
\end{array}  \right]
\label{defTransform}
\end{equation}

This is a linear transformation, with 
\begin{equation}
\setlength{\arraycolsep}{4pt}
\renewcommand{\arraystretch}{1.8}
\mathsfbi{T} = \left[
\begin{array}{cc}
   \displaystyle \frac{1}{2} & \displaystyle\frac{\sqrt{3}}{2} \\
   1 & 0 \\
\end{array}  \right]
\label{defT}
\end{equation}
being the transformation matrix. 

With the assumption of no ternary diffusion and $D_\mathrm{e}=D_\mathrm{o}\equiv D$, \cite{ruschak1972spontaneous} first proved that the diffusion path in the ternary phase diagram is a straight line. By definition, dilution curve is also a straight line. Because linear transformation transforms straight lines to straight lines, diffusion path AB and dilution curve DM in figure \ref{fig:Trans}($c$) are still straight lines.

On the dilution curve DM in figure \ref{fig:Trans}($a$), ratio of water weight fraction to ethanol weight fraction is kept constant, that is:
\begin{equation}
w_\mathrm{w}/w_\mathrm{e}=c
\label{eq:DiluCurveDef}
\end{equation}
where $c$ is a positive constant. 

In the cartesian coordinate, $w_\mathrm{e}$ is the X-coordinate and $w_\mathrm{o}$ is the Y-coordinate. For easier notation, let $x=w_\mathrm{e}$ and $y=w_\mathrm{o}$, then $w_\mathrm{w}=1-x-y$, and 

\begin{equation}
\left. \begin{array}{l}  
\quad\quad 0\leq x \leq 1\\
\quad\quad 0\leq y \leq 1\\
0\leq 1-x-y \leq 1
\end{array}\right\}
\label{eq:Condition}
\end{equation}

Eq.(\ref{eq:DiluCurveDef}) can be transformed to 
\begin{equation}
y=1-(1+c)x
\label{eq:DiluCurveCartesian}
\end{equation}

This is a straight line passing through (0,1) in figure \ref{fig:Trans}($c$). Its slope satisfies $-\infty\leq -(1+c) \leq -1$. This is also why the phase diagram in figure \ref{fig:Trans}($c$) is a right triangle.

\section{Influence of non-equal diffusivities}\label{appDiff}

Normally, the diffusivities of water and oil in the mixture are not equal. Here we briefly discuss the influence of non-equal diffusivities of water and oil $D_\mathrm{w}\neq D_\mathrm{o}$ on the prefactor $s$. 

With the assumption that ternary diffusion effects are neglected, and in the meantime ignore the advection term in the advection-diffusion equations, the transport equations for water and oil become \citep{ruschak1972spontaneous}:

\begin{equation}
\frac{\partial w_\mathrm{w}}{\partial t}=D_\mathrm{w}\frac{\partial^2 w_\mathrm{w}}{\partial x^2}
\label{eq:DiffControl1}
\end{equation}
\begin{equation}
\frac{\partial w_\mathrm{o}}{\partial t}=D_\mathrm{o}\frac{\partial^2 w_\mathrm{o}}{\partial x^2}
\label{eq:DiffControl2}
\end{equation}

Solutions of Eqs.(\ref{eq:DiffControl1})\&(\ref{eq:DiffControl2}) have the form:

\begin{equation}
w_\mathrm{w}=\frac{w_\mathrm{w,A}+w_\mathrm{w,B}}{2}+\frac{w_\mathrm{w,A}-w_\mathrm{w,B}}{2} \mathrm{erf} \frac{x}{2tD_\mathrm{w}}
\label{eq:ww}
\end{equation}
\begin{equation}
w_\mathrm{o}=\frac{w_\mathrm{o,A}+w_\mathrm{o,B}}{2}+\frac{w_\mathrm{o,A}-w_\mathrm{o,B}}{2} \mathrm{erf} \frac{x}{2tD_\mathrm{o}}
\label{eq:wo}
\end{equation}
where $x$ and $t$ are the distance (as defined in figure \ref{fig:4model}($d$)) and time, respectively. $w_\mathrm{w,A}$ and $w_\mathrm{w,B}$ are the water weight fraction of solutions A and B, $w_\mathrm{o,A}$ and $w_\mathrm{o,B}$ are the oil weight fractions of solutions A and B. With $-\infty<x<+\infty$, and $w_\mathrm{e}=1-w_\mathrm{w}-w_\mathrm{o}$,  the diffusion path can be computed.

\begin{figure}
	\centering
	\includegraphics[width=0.5\linewidth]{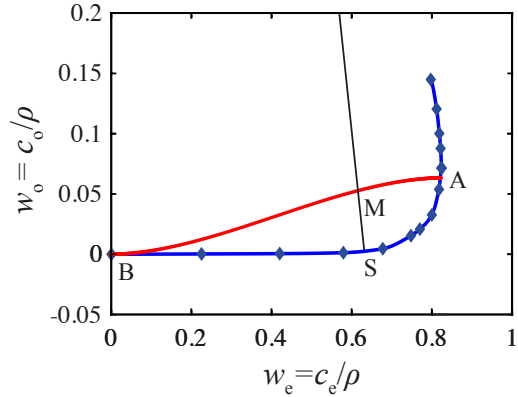}
\caption{Diffusion path (red curve) for the case when $D_\mathrm{w}=\SI{0.84e-9}{m^2/s}$ and $D_\mathrm{o}=\SI{0.43e-9}{m^2/s}$, with initial conditions $w_\mathrm{o,A}=0.063$, $w_\mathrm{w,A}=0.119$. Notice now the diffusion path (red curve AB) changes to a curve rather than a straight line as shown in figure \ref{fig:4model}($b$). The blue diamonds are the measured data on the binodal, and the blue curve is the binodal. The black line is the dilution curve.}
\label{fig:S3}
\end{figure}

Using the same diffusivity of water $D_\mathrm{w}=\SI{0.84e-9}{m^2/s}$ and take the oil diffusivity $D_\mathrm{o}=\SI{0.43e-9}{m^2/s}$ as calculated by \cite{perkins1969molecular}, we obtain the diffusion path for initial conditions $w_\mathrm{o,A}=0.063$, $w_\mathrm{w,A}=0.119$, as shown in figure \ref{fig:S3}. The diffusion path is no longer a straight line, but curved. $s$ is calculated to be $s=\SI{1.2e-2}{}$ and \SI{3e-3}{} for the two initial conditions $w_\mathrm{o,A}=0.063$ and 0.017, all less than \SI{4}{\%} change as compared to the case when assuming $D_\mathrm{w}=D_\mathrm{o}$.


\end{document}